\documentstyle[prl,tighten,aps,epsf,amstex,multicol,bbm,times]{revtex}
\begin{document}

\draft

\newcommand{\proofend}{\hfill\rule[0pt]{.2cm}{.2cm}}

\title{Distilling Gaussian states with Gaussian operations is impossible}

\author{J.~Eisert, S.~Scheel, and M.B.~Plenio}

\address{QOLS,
Blackett Laboratory, Imperial College of Science, Technology 
and 
Medicine, London, SW7 2BZ, UK}

\date{\today}

\maketitle

\begin{abstract}    
We will show that no distillation protocol for Gaussian
quantum states exists that relies on (i) arbitrary
local unitary  operations that preserve the Gaussian character of
the state and (ii) homodyne detection together with classical
communication and postprocessing by means of 
local Gaussian unitary operations 
on two symmetric identically prepared
copies. This analysis shows that unlike
the finite-dimensional case, where entanglement can be 
distilled in an iterative protocol using two copies at a time,
there is no such procedure in the case of continuous variables
for Gaussian initial
states and the above Gaussian operations. 
The ramifications for the distribution of Gaussian states
over large distances will be outlined. 
We will also comment on the generality of the approach and sketch
the most general form of a Gaussian local operation 
with classical communication in a bi-partite setting.
 \end{abstract}


\maketitle

\begin{multicols}{2}
\narrowtext

In most practical implementations of information processing
devices sophisticated methods are necessary in order to
preserve the coherence of 
the involved quantum states.
Even the mere preparation of an entangled state
of spatially distributed quantum systems 
requires 
such techniques: once prepared
locally and then distributed, an entangled state
will to some extent deteriorate from 
a highly entangled
state to a less correlated state through
the process of decoherence. This process can
quite obviously not be entirely avoided.
However, one may prepare
and distribute several identical entangled states,
and then apply appropriate partly measuring
local quantum operations and classical communication 
to obtain states that
are similar to the highly entangled original state. 
This is only possible at the expense that one 
has fewer identically prepared systems
or copies at hand, but
this is a small price to pay.
Appropriately indeed, this process has been given the 
name distillation \cite{Distill}, 
as fewer more
highly entangled states are 'distilled' from a supply 
of many less entangled states. It was one of the
major successes of the field of quantum information science
to realize that for two-level systems such a distillation
procedure may be performed on only two copies at a time,
and it requires only two steps: (i) a local unitary operation,
and (ii) a local measurement, together with the classical
communication about the measurement outcome. Based on
the measurement outcome further local unitary
operations are then implemented.

Such distillation protocols may also be 
of crucial importance
in the infinite-dimensional setting. 
Quantum information science over
continuous variables has seen an enormous
progress recently, both in theory and experiment,
mostly involving Gaussian states of field modes
in a quantum optical setting \cite{Other,LOG,AllDist}. 
Quite naturally, one should expect that a similar 
distillation procedure also works for Gaussian states in
the infinite-dimensional case, also 
under the preservation of 
the Gaussian character of the state.
If one transmits two 
pure two-mode squeezed 
Gaussian states through lossy 
optical systems such as fibers, the corresponding modes being
from now on labeled $A1$, $A2$, $B1$, and $B2$,
one obtains two identical copies of
less entangled symmetric states \cite{Scheel}. 
A feasible distillation protocol preserving
the Gaussian character may
consist of the subsequent steps (see Fig.\ 1):
 
(i)     Application of any local Gaussian unitary operation. 
	That is, one may implement
	any unitary operations $U_A$ and $U_B$ on both
	$A1$ and $A2$ on one hand and $B1$ and $B2$ on
	the other hand
	corresponding to symplectic transformations \cite{Symplectic}
	$S_{A},S_{B}\in Sp(4,{\mathbb{R}})$  \cite{Displace}.
	This set 
	includes all two-mode and one-mode squeezings,
        mixing at beam splitters and
        phase shifts. To specify these operations
	20 real parameters are necessary.
	Note that we do not require
	both parties to realize the same transformation.

(ii) A homodyne
measurement on the modes $A2$ and $B2$.
The parties communicate classically about the outcome
of the measurement, and may postprocess the states of
modes $A1$ and $B1$
with unitary Gaussian operations.

\begin{figure}[ht]
\centerline{
       \epsfxsize=4.2cm
       \epsfbox{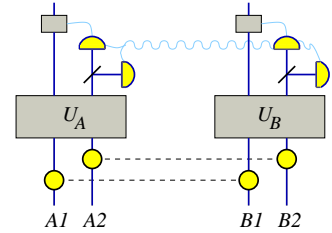}}
\smallskip

\caption{The class of considered feasible
distillation protocols.}
\end{figure}

The main result of this Letter is that 
very much as a surprise, none of these protocols
amounts to a distillation protocol.
No matter how ingeniusely 
the local unitary operation is chosen,
the degree of entanglement can not be increased.
The optimal procedure is simply to do nothing at all, 
which means that at least no entanglement is lost 
\cite{Butdist}. 
The degree of entanglement will be measured in terms of
the log-negativity, which is defined as
$E_{N}(\rho)= \log_2 \|\rho^{T_{A}}\|$ for a state
$\rho$, where
$\|.\|$ denotes the trace norm, and $\rho^{T_{A}}$
is the partial transpose of $\rho$.
The negativity has been
shown to be an entanglement measure in the sense that it
is non-increasing 
on average under local operations with classical 
communication \cite{Nega}, and is to date 
the only known feasible measure of entanglement 
for Gaussian states. For pure (and for symmetric
mixed) Gaussian states it is related to the
degree of squeezing in a monotone way
(see, e.g., \cite{WolfSqueeze}). 
This means that as a corollary of the main result, 
it follows that with Gaussian operations as
specified above one cannot transform two identically prepared
two-mode squeezed vacua into a single two-mode squeezed vacuum
state with a higher degree of squeezing.

We will start by fixing the notation.
Gaussian states \cite{covar} of an $n$-mode system
are completely characterized by their first and second
moments. The first moments are the expectation 
values of the canonical coordinates.
The second moments can be collected in 
the real symmetric
covariance matrix
$\Gamma\!\in\!C(2n)\!\subset\! M({2n},{\mathbb{R}})$,
where $M({2n},{\mathbb{R}})$ denotes the 
set of real $2n\times 2n$-matrices, and $C(2n)$ the subset
of matrices obeying the Heisenberg uncertainty principle \cite{covar}.
The linear transformations from 
one set of canonical coordinates to another
which preserve the
canonical commutation relations
form the group of real linear
symplectic transformations $Sp(2n ,\mathbb{R})$ \cite{Symplectic}.
A symplectic
transformation $S$
changes the covariance matrix
according to $\Gamma\longmapsto S \Gamma S^T$,
while
states undergo a unitary operation 
$\rho\longmapsto U\rho U^\dagger$.
The $n\!=\!4$ 
modes $A1$, $A2$, $B1$, $B2$
will be equipped with the canonical operators 
$(X_{A1},P_{A1}, ... , X_{B2}, P_{B2})$.
To make the notation more transparent, both
tensor products and direct sums will carry a 
label indicating the underlying split, meaning
either $A,B$ or $1,2$. 
We state the main result of this Letter 
in form of a theorem:

\smallskip\noindent
{\bf Theorem. --} {\it Let $\rho\otimes \rho$ be
two identically prepared symmetric Gaussian states
of two-mode systems consisting of the parts
$A1$, $A2$, $B1$, and $B2$, respectively, each of
which having the covariance matrix
\begin{equation}
    \Gamma^{{(0)}}=\left(
    \begin{array}{cccc}
    a & 0 & c & 0\\
    0 & a & 0 & -c\\
    c & 0 & a & 0 \\
    0 & -c & 0 & a \\
    \end{array}\right),\,\, a\geq 1,\,\, 0 \leq c\leq (a^{2}-1)^{1/2},
\end{equation}    
let $S_{A},S_{B} \in Sp(4,{\mathbb{R}})$
be any symplectic
transformations
with associated unitaries $U_{A}$ and $U_{B}$, and let
\begin{equation}
\rho'= (U_{A}\otimes_{A,B} U_{B})(\rho\otimes_{1,2} \rho) 
(U_{A}\otimes_{A,B} U_{B})^\dagger.
\end{equation}
Then any state $\rho''$ 
that is obtained from $\rho'$
via a selective 
homodyne measurement on
systems $A2$ and $B2$ satisfies
	$E_N(\rho'')\leq E_N(\rho)$,
that is, the degree of entanglement can only decrease.}

The proof of this statement
will turn out to be technically involved,
and while the statement itself is concerned with 
practical quantum optics,
the techniques used in the proof will be mostly
taken from matrix analysis \cite{Horn}. 
In order to give the general argument more structure, 
the proof is 
split into several lemmata. The entire proof
will be formulated in terms of covariance matrices,
rather than in terms of the states.

The log-negativity of a state $\sigma$ 
of a two-mode system can be 
easily expressed in terms of the entries
of the associated covariance matrix
$\gamma\in C(4)$. The latter 
can be partitioned in block form according to 
\begin{equation}
    \gamma=\left(
    \begin{array}{cc}
	\gamma_{A} & \gamma_{C}\\
	\gamma_C^{T} & \gamma_{B}\\
    \end{array}
    \right), \,\,\, \gamma_{A},\gamma_{B},
\gamma_{C}\in M(2,R). \label{block}
\end{equation}
The log-negativity $E_{N}(\sigma)$
is then given by \cite{Nega}
\begin{eqnarray}\label{logne}
    E_{N}(\sigma)=\left\{
    \begin{array}{cc}
	-(\log \circ f)(\gamma)/2,
	& \text{if }
	f(\gamma)<1,\\
	0& \text{otherwise.}
	\end{array}
    \right.
\end{eqnarray}
where the function 
$f:C(4)\longrightarrow
{\mathbb{R}}^{+}$  is defined as
\begin{eqnarray}
    f(\gamma)&=&((\text{det}[\gamma_{A}]+ 
    \text{det}[\gamma_{B}])/2-\text{det}[\gamma_{C}])\nonumber\\
    &-&
    \bigl[(    (\text{det}[\gamma_{A}]+ \text{det}[\gamma_{B}])/2 
    -\text{det}[\gamma_{C}])^{2}-\text{det}[\gamma]\bigr]^{1/2}.
\end{eqnarray}
The covariance matrix associated with the Gaussian state
$\rho'$ in the Theorem will be denoted as $\Gamma'\in C(8)$.
For any $S_A,S_B \in Sp(4,{\mathbb{R}})$ this 
covariance matrix of the modes $A1$, $A2$, $B1$, and $B2$
becomes
$\Gamma':=
(S_A\oplus_{A,B} S_B) (\Gamma^{(0)}\oplus_{1,2} \Gamma^{(0)})
(S_A\oplus_{A,B} S_B)^T$
The first step is to relate the covariance matrix
$\Gamma''$ associated with the state after 
the measurement to a Schur complement \cite{Horn}. 
This Schur complement
structure is a general feature of Gaussian operations and
will be further discussed at the end of the letter.

\smallskip\noindent
{\bf Lemma 1. --} {\it Let $\Gamma'\in C(8)$
be a covariance matrix of systems $A1$,
$A2$, $B1$, and $B2$ associated with a state
$\rho'$, which can be written in block form 
as
\begin{equation}\label{p1}
\Gamma'=\left(
    \begin{array}{cc}
	C_{1} & C_{3}\\
	C_{3}^{T} & C_{2}\\
    \end{array}
    \right),
\end{equation}
where $C_1,C_2, C_3 \in M(4,{\mathbb{R}})$. 
The covariance matrix of the state 
that is obtained by a projection in $A2$ and $B2$ 
on the pure Gaussian state with covariance matrix
$D_d:= \text{diag}
(1/d,d,1/d,d)\in C(4)$, 
$d>0$, 
is then given by
\begin{equation}\label{gamma2strich}
    M_d=
        C_{1}-C_{3}(C_{2}+ D_d^2 )^{-1}C_{3}^{T}.
\end{equation}
}
\indent {\it Proof.} This statement can be most conveniently
be shown in terms of the characteristic function $\chi$ \cite{covar}.
By employing the Weyl (displacement) operator,
the state $\rho'$
associated with the covariance matrix
$\Gamma'$ can be
written in terms 
of the characteristic function according to
	$\rho'= (1/\pi^4) 
	\int d^8 \xi \, W(-\xi) \chi(\xi)$ (see, e.g., Ref.\ 
\cite{VogelWelsch}).
The projection corresponds on the level
of the characteristic function therefore to
an incomplete Gaussian integration.
The characteristic function 
associated with 
the modes $A1$ and $B1$ can then be written as
\begin{eqnarray}
	\chi(\xi_1,\dots , \xi_4)
	&= &
	\int \frac{d\xi_5 \dots d\xi_8}{\pi^2}
	e^{
	- \xi^T  \Gamma' \xi/2}
  	e^{-\frac{1}{2 d^2} (\xi_5^2+\xi_7^2) - \frac{d^2}{2} (\xi_6^2 + \xi_8^2)}\nonumber
	\\
	&=&
	| C_{2}+D_d^2|^{-1/2}
	e^{ -(\xi_1,\dots ,\xi_4)
  	 \Gamma''  (\xi_1,\dots ,\xi_4)^{T}/2},
\end{eqnarray}
with $M_d$ 
defined as in Eq.~(\ref{gamma2strich}).
\proofend

Hence, the resulting covariance matrix is given by
the Schur complement $C_{1}-C_{3}(C_{2}+D_d^2)^{-1}C_{3}^{T}$ 
of the matrix
\begin{equation}\label{gammadprime}
	\Gamma_d':=
	\left(
    \begin{array}{cc}
	C_{1} & C_{3}\\
	C_{3}^{T} & C_{2}+D_d^2 \\
    \end{array}
    \right)
\end{equation}
with respect to the leading principal submatrix $C_1$.
The additional matrix $D_d^2$
originates from the projection in the modes $A2$ and $B2$. 
Note that although this Lemma has been formulated 
in terms of the projection on a certain class of pure Gaussian states,
it applies to the projection on {\it any}
 pure Gaussian state
in the modes $A2$ and $B2$: the projection on any
other pure Gaussian state can be realized by an appropriate
choice of the symplectic transformations $S_{A}$ and $S_{B}$.
Ideal homodyne detections can now be formulated
as projections on `infinitely squeezed' pure Gaussian states 
\cite{VogelWelsch}.
The central feature is that  
the initial first moments do not affect the form
of the covariance matrix after the measurement.
Lemma 2 gives the form of the resulting covariance matrix in
case of a homodyne detection in modes $A2$ and $B2$. In the 
limit $d\rightarrow 0$ the  matrix $D_d$
gives rise to a projection operator, and the
inverse becomes a Moore Penrose inverse (MP)  \cite{Horn}:

\smallskip\noindent
{\bf Lemma 2. --} {\it In the notation of Lemma 1, the covariance
matrix of modes $A1$ and $B1$ after a selective homodyne measurement
in modes $A2$ and $B2$ is given by
\begin{equation}\label{sq}
    \Gamma'':=\lim_{d\rightarrow 0} \Gamma_{d}=
        C_{1}-C_{3}(\pi C_{2} \pi)^{\text{MP}}C_{3}^{T},
\end{equation}
where 
$\pi=\text{diag}(1,0,1,0)$.
}

Equipped with these preparatory considerations, 
we will now turn to the core of the proof.
In order to be able to evaluate the logarithmic negativity
according to Eq.\ (\ref{logne}),
one needs to know the values of the invariants under
local symplectic transformations, i.e., the determinants
of four submatrices. To find an expression for all
these determinants is however a quite difficult task.
Instead, we will later make use of an upper bound of
the logarithmic negativity that only involves
determinants of principal submatrices \cite{Horn} 
of $\Gamma''$.

\smallskip\noindent
{\bf Lemma 3. --} {\it Let $\Gamma''\in C(4)$ be
defined as in Lemma 2. Then, independent of $S_{A},S_{B}\in Sp(4,{\mathbbm{R}})$,
\begin{equation}
\text{det}[\Gamma'']=
\text{det}[\Gamma^{(0)}]=(a^{2}-c^{2})^{2}.
\end{equation}
}
{\it Proof.} According to Lemma 2, $\Gamma''$ is given by
$\Gamma''= \lim_{d\rightarrow 0} M_d$. The
Schur complement of the matrix $\Gamma'_d$ as
defined in Eq.\
(\ref{gammadprime})
is related to
 $\Gamma'_d$ and one of its principal submatrices
via the congruence
\begin{equation}\nonumber
	\left(
    \begin{array}{cc}
	\openone_4 & X\\
	0 & \openone_4
\end{array}
    \right)
	\Gamma'_d	\left(
    \begin{array}{cc}
	\openone_4 & 0\\
	X^T & \openone_4
\end{array}
    \right)
	= \left(
    \begin{array}{cc}
    \Gamma_d & 0\\
    0 & C_{2}+D_d^2
    \end{array}
    \right),
\end{equation}
where $X:= -C_3(C_2+D_d^2)^{-1}$. 
Hence, according to the determinant multiplication
theorem  we obtain 
	$\text{det}[\Gamma'_d] =\text{det}[\Gamma_d] \, \text{det}[C_2+D_d^2]$,
which yields in the limit $d\rightarrow 0$
\begin{equation}\nonumber
\text{det}[\Gamma''] 
= \text{det}[P \Gamma'_d P+ (\openone_8-P) ]/ 
\text{det}[Q \Gamma'_d  Q+ (\openone_8 -Q)],
\end{equation}
where the projections $P$ and $Q$ are defined as
$
P:=\text{diag}(1,1,1,1,0,1,0,1)$ and
$Q:= \text{diag}(0,0,0,0,0,1,0,1)$.
With these tools, it is feasible to directly
prove the statement of Lemma 3 by
parameterizing $S_A,S_B \in Sp(4,{\mathbbm{R}})$.
Every $S\in Sp(4,{\mathbbm{R}})$ can be
written as a product $S=V D W$, where
$V,W\in Sp(4,{\mathbbm{R}})\cap SO(4)$, and
$D:=\text{diag}(d_1,1/d_1, d_2,1/d_2)$ with
$d_1,d_2\in {\mathbbm{R}}$ \cite{Symplectic}.\proofend

%

\smallskip\noindent
{\bf Lemma 4. --} {\it Let $\Gamma''
\in C(4)$ 
be defined as in Lemma 2, and let $\Gamma_A''$ and $\Gamma_B''$ be
the principal submatrices belonging to mode $A1$ and $B1$.
Then, for all
$S_A,S_B\in Sp(4,{\mathbbm{R}})$,
\begin{eqnarray}
	\text{det}[\Gamma_A''] \leq \text{det}[\Gamma^{(0)}_{A}] =
	a^2,\,\,
	\text{det}[\Gamma_B''] \leq \text{det}[\Gamma^{(0)}_{B}] =
	a^2.
\end{eqnarray}
}
{\it Proof.} ${\Gamma''}$ is defined as the covariance matrix
corresponding to modes $A1$ and $B1$
after the projective measurements in both
$A2$ and $B2$. Let us assume that one first performs the
projective measurement in $A2$, leading to a 
the covariance matrix $N_{A}\in C(2)$ 
of the reduced state of $A1$. The covariance
matrix $\Gamma_A''$ after the projection in $B2$ is 
then obtained as a Schur complement. In particular,
$\Gamma_A''$ can be written as $\Gamma_A''= N_{A}- R$, where 
$R\in M(2,{\mathbbm{R}})$
is a real symmetric positive matrix. 
Hence, as $\Gamma_A''$ and $N_{A}$ are also positive,
$\text{det}[\Gamma_A'']\leq \text{det}[N_{A}]$
\cite{Horn}.
In other words, one obtains an upper 
bound for $\text{det}[\Gamma_A'']$
when considering only a projective measurement in $A1$. The statement
of Lemma 4 follows from Lemma 3 in the special 
case that $c=0$: one can after a few steps 
conclude that then
$\text{det}[N_{A}]=a^{2}$, independent of $S_{A},S_{B}\in 
Sp(4,{\mathbbm{R}})$. The same reasoning applies to $\Gamma_B''$.\proofend

The most important step is now an appropriate
upper bound of the log-negativity of the resulting state. The actual bound
might appear somewhat arbitrary, but it will turn out
that it is exactly the tool that we need in the
last step of the proof. 

\smallskip\noindent
{\bf Lemma 5. --} {\it 
Let $\gamma\in C(4)$, partitioned as in Eq.\ (\ref{block}). Then
\begin{eqnarray}
	f(\gamma)\geq g(\gamma) &:=&
	\bigl[ \left((\text{det}[\gamma_A] + \text{det}[\gamma_B])/2
	\right)^{1/2}\\
	&-& (
	(\text{det}[\gamma_A] + \text{det}[\gamma_B])/2- \text{det}[\gamma]^{1/2}
	)^{1/2}
	\bigr]^2.\nonumber
\end{eqnarray}
}
{\it Proof.} $g(\gamma)$ can be expressed in terms of $f$ as 
$g(\gamma)= f(\gamma')$, 
\begin{equation}\nonumber
    \gamma'=\left(
    \begin{array}{cc}
	\gamma_{A}'& \gamma_{C}'\\
	{\gamma_{C}'}^{T} & \gamma_{B}'
	\end{array}
    \right),\,\,\,
	\gamma'_A= \gamma'_B = a' \openone_2, \,\,\,
	\gamma'_C=  
	\left(
	\begin{array}{cc}
	c' & 0 \\
	0 & -c'\\
	\end{array}
	\right),
\end{equation}
where $a':= ((\text{det}[\gamma_A]^2 + \text{det}[\gamma_B]^2)/2)^{1/2}$ and
$c':= ( {a'}^2 - \text{det}[\gamma]^{1/2} )^{1/2}$. Hence, one 
has to prove
that $f(\gamma')\leq f(\gamma)$. Firstly,
note that $\text{det}[\gamma]=\text{det}[\gamma']$. Secondly, 
$(\text{det}[\gamma_A]+\text{det}[\gamma_B])/2={a'}^2$.
Therefore, it remains to be shown that 
${c'}^2\geq | \text{det}[\gamma_C]|$.
This inequality is equivalent with
    $\left[
    (\text{det}[\gamma_{A}]+
    \text{det}[\gamma_{B}])/2
    -|\text{det}[ \gamma_{C}]\,|
    \right]^{2}- \text{det}[\gamma]\geq 0$,
which is a valid inequality, as $\gamma\in C(4)$.
\proofend

{\it Proof of the Theorem.}
Let ${\Gamma''}\in C(4)$ be the matrix defined as in Lemma 2. The
log-negativity of the corresponding state of modes $A1$ and $B1$
is given by $-(\log \circ f)({\Gamma''})$,  if the final state is entangled at
all, as we will assume from now on.
Lemma 5 yields the bound $f({\Gamma''})\geq g({\Gamma''})$. In $g({\Gamma''})$, however,
only the determinants of the principal submatrices are needed,
bounds of which are available by virtue of 
Lemma 3 and 4. The function $h:[y,\infty)\rightarrow{\mathbbm{R}}^{+}$ 
with
$h(x)= (x^{1/2}- ( x - y)^{1/2})^{2}$, $y>0$,
is a strictly monotone decreasing function of $x$. 
Therefore, using Lemma 3 and 4 
one can conclude that $g({\Gamma''}) \geq g(\Gamma^{(0)})$.
Moreover, $g(\Gamma^{(0)})= f(\Gamma^{(0)})$, due to
the special form of $\Gamma^{(0)}$, as can be easily verified.
Hence,
    $f({\Gamma''})\geq g(\Gamma^{(0)})= f(\Gamma^{(0)})$,
which leads to
$-(\log\circ f)({\Gamma''})\leq - (\log\circ f) 
(\Gamma^{(0)})$. This is finally
the desired result: it means that the degree of entanglement can only
decrease. \proofend

We will finally comment on the generality of the approach.
A general Gaussian operation
is a quantum operation that maps all Gaussian states
on Gaussian states \cite{LOG}. 
Any general Gaussian local 
operation with classical communication
(LOCCG) -- 
trace-preserving or non-trace-preserving
--
can be decomposed
into the subsequent steps:
(i) Appending locally additional
modes that have been prepared in a 
Gaussian state \cite{LOG}.
(ii) Application of 
any local unitary Gaussian operation on both the original and 
the additional system. These comprise 
operations corresponding
to symplectic transformations and displacements in phase space.
(iii) Projections
on pure Gaussian states or ideal homodyne detections,
which give rise to Schur complements on the level of covariance
matrices as described above, together with
the classical communication about the outcome (real numbers in 
case of homodyne detection, bits in case of dichotomic
measurements including the projection 
on a pure Gaussian state in one outcome), (iv)
mixing, such that the resulting state is Gaussian, 
and (v) a partial trace, which corresponds to considering a certain
principal submatrix of the covariance matrix 
only \cite{Note}. 
The proof is therefore restrictive in the sense that 
only two copies at a time are considered, other projections on
Gaussian states are excluded, and no additional modes 
are allowed for. 
The statement of the present paper proves that iterative
protocols in strict analogy to the corresponding methods
in finite dimensional settings certainly
do not work. Indeed, the findings strongly suggest that Gaussian states 
cannot be distilled at all 
with Gaussian operations. 
Then (less feasible)
non-linear physical effects \cite{Barry}
would have to be made use of in 
order to distill from a supply of Gaussian 
two-mode states  \cite{Nonlin}. Such techniques would then also
be necessary for the realistic
implementation of quantum repeaters \cite{Briegel}
for continuous-variable systems when it comes to 
the distribution of highly entangled Gaussian states over
large distances. 

We would like to thank K.\ Audenaert,  J.\ Fiur\'a\v{s}ek,
P.\ van Loock, C.\ Silberhorn, 
G.\ Giedke, J.I.\ Cirac,
S.D.\ Bartlett, 
B.C.\ Sanders, D.-G.\ Welsch, and  N.\  Cerf
for discussions. This work has been supported by the 
European Union
(EQUIP, QUEST) 
and the A.-v.-Humboldt-Foundation.

\end{multicols}


\begin{thebibliography}{99}
\bibitem{Displace}
	Both parties may also apply
	displacements in phase space, which can always be implemented
	locally.
\bibitem{Butdist}
	In contrast, it has been shown that 
	from any bi-partite
	Gaussian state with
	a non-positive partial transpose 
	singlets can be distilled in principle
	using general quantum operations (obviously under 
	loss of the Gaussian character of the state) \cite{AllDist}.
\bibitem{covar}
	In terms of the $2n$ canonical coordinates
	$
	(O_{1},\ldots,O_{2n}):=  (X_{1},P_{1},\ldots,X_{n},P_{n})$
	the entries of the covariance matrix are
	given by
        $\Gamma_{jk}
        =
        2 \text{tr}\left[
        \rho \left(O_j-\langle O_j\rangle_{\rho} \right) 
        \left(O_k-\langle O_k\rangle_{\rho}  \right)
        \right]  - i \Sigma_{jk}$,
	 $j,k=1,\ldots,2n$,
	where the 
	skew-symmetric 
	$2n\times 2n$-matrix
	$\Sigma$ 
	is defined as 
	$[O_j, O_k]=i\Sigma_{jk}$. Any symmetric
	matrix $\Gamma\in M(2n,{\mathbb{R}})$ satisfying
	the Heisenberg uncertainty principle 
	$\Gamma \geq i \Sigma$
	is a 
	covariance matrix.
	The defining property of Gaussian states is that
	their characteristic function 
	$\chi({\mathbf{\xi}})= \text{tr}[W({\mathbf{\xi}}) \rho]$
	is a Gaussian function 
	in phase space, where
	$W({\mathbf{\xi}}):=\text{exp}
	( -i \sum_{j,k=1}^{2n} \xi_j \Sigma_{jk} O_k)$
	denotes the Weyl operator. The Wigner function, as often employed in this 
	context,
	is given by the Fourier transform of the characteristic function.
\bibitem{Note}
	Note that Gaussian operations including projections on
	pure Gaussian states can also 
	be implemented by means of 
	symplectic transformations and homodyne measurements.
\end{thebibliography}
\end{document}